\documentclass[aps,prl,reprint,secnumarabic,twocolumn,superscriptaddress,tshowpacs,nobibnotes,floatfixol,byrevtex]{revtex4-1}

\usepackage{pst-grad,epsfig,calc,subfigure}
\usepackage{amsmath,amssymb}
\usepackage{multirow}
\usepackage{xcolor}
\usepackage{xr}
\usepackage{epsfig}
\usepackage{graphicx}
\usepackage{float}
\usepackage{tabularx}
\usepackage{blindtext}
\usepackage{color}
\usepackage[colorlinks=true,allcolors=blue,citecolor=blue]{hyperref}
\usepackage[nameinlink]{cleveref}
\usepackage{natbib}
\usepackage{smartref}
\usepackage[utf8]{inputenc}
\usepackage{color,soul}

\begin{document}

\title{$\sqrt{2}$$\times$$\sqrt{2}R45^\circ$ surface reconstruction and electronic structure of BaSnO$_3$ film}

\author{Shoresh Soltani}
\altaffiliation{Present address: SUPA, School of Physics and Astronomy, University of St. Andrews, St. Andrews KY16 9SS, United Kingdom}
\affiliation{Institute of Physics and Applied Physics, Yonsei University, Seoul 03722, Republic of Korea}
\affiliation{Center for Correlated Electron Systems, Institute for Basic Science (IBS), Seoul 08826, Republic of Korea}

\author{Sungyun Hong}
\affiliation{Department of Physics and Astronomy, Seoul National University (SNU), Seoul 08826, Republic of Korea}

\author{Bongju Kim}
\affiliation{Center for Correlated Electron Systems, Institute for Basic Science (IBS), Seoul 08826, Republic of Korea}

\author{Donghan Kim}
\affiliation{Center for Correlated Electron Systems, Institute for Basic Science (IBS), Seoul 08826, Republic of Korea}
\affiliation{Department of Physics and Astronomy, Seoul National University (SNU), Seoul 08826, Republic of Korea}

\author{Jong Keun Jung}
\affiliation{Center for Correlated Electron Systems, Institute for Basic Science (IBS), Seoul 08826, Republic of Korea}
\affiliation{Department of Physics and Astronomy, Seoul National University (SNU), Seoul 08826, Republic of Korea}

\author{Byungmin Sohn}
\affiliation{Center for Correlated Electron Systems, Institute for Basic Science (IBS), Seoul 08826, Republic of Korea}
\affiliation{Department of Physics and Astronomy, Seoul National University (SNU), Seoul 08826, Republic of Korea}

\author{Tae Won Noh}
\affiliation{Center for Correlated Electron Systems, Institute for Basic Science (IBS), Seoul 08826, Republic of Korea}
\affiliation{Department of Physics and Astronomy, Seoul National University (SNU), Seoul 08826, Republic of Korea}

\author{Kookrin Char}
\affiliation{Department of Physics and Astronomy, Seoul National University (SNU), Seoul 08826, Republic of Korea}

\author{Changyoung Kim}
\email{changyoung@snu.ac.kr}
\affiliation{Center for Correlated Electron Systems, Institute for Basic Science (IBS), Seoul 08826, Republic of Korea}
\affiliation{Department of Physics and Astronomy, Seoul National University (SNU), Seoul 08826, Republic of Korea}

\date{\today}

\begin{abstract}

We studied surface and electronic structures of barium stannate (BaSnO$_3$) thin-film by low energy electron diffraction (LEED), and angle-resolved photoemission spectroscopy (ARPES) techniques. BaSnO$_3$/Ba$_{0.96}$La$_{0.04}$SnO$_3$/SrTiO$_3$ (10 nm/100 nm/0.5 mm) samples were grown using pulsed-laser deposition (PLD) method and were \emph{ex-situ} transferred from PLD chamber to ultra-high vacuum (UHV) chambers for annealing, LEED and ARPES studies. UHV annealing starting from 300$^{\circ}$C up to 550$^{\circ}$C, followed by LEED and ARPES measurements show 1$\times$1 surfaces with non-dispersive energy-momentum bands. The 1$\times$1 surface reconstructs into a $\sqrt{2}$$\times$$\sqrt{2}R45^\circ$ one at the annealing temperature of 700$^{\circ}$C where the ARPES data shows clear dispersive bands with valence band maximum located around 3.3 eV below Fermi level. While the $\sqrt{2}$$\times$$\sqrt{2}R45^\circ$ surface reconstruction is stable under further UHV annealing, it is reversed to 1$\times$1 surface by annealing the sample in 400 mTorr oxygen at 600$^{\circ}$C. Another UHV annealing at 600$^{\circ}$C followed by LEED and ARPES measurements, suggests that LEED $\sqrt{2}$$\times$$\sqrt{2}R45^\circ$ surface reconstruction and ARPES dispersive bands are reproduced. Our results provide a better picture of electronic structure of BaSnO$_3$ surface and are suggestive of role of oxygen vacancies in the reversible  $\sqrt{2}$$\times$$\sqrt{2}R45^\circ$ surface reconstruction.

\begin{description}
\item[PACS numbers]

\end{description}
\end{abstract}

\keywords{Suggested keywords}
\maketitle

\section{Introduction}
Barium stannate (BaSnO$_3$) has recently been widely studied for its interesting electronic and optical properties. As a wide band gap perovskite oxide, it has simultaneously shown high room-temperature mobility and optical transparency, making it interesting for applications ranging from optoelectronics and solar energy to functional devices\cite{Shan,Ismail-Beigi,SMXing,WJ LEE,Gogova,YajunZhang}. Its high room temperature mobility suggests for power electronics applications due to its low power consumption\cite{XLuo,Prakash}. The pioneering work of Kim \emph{et al} \cite{HJKim} in discovering high room temperature mobility of 300 cm$^{2}$V$^{-1}$s$^{-1}$ in La-doped BaSnO$_3$ (BLSO) suggested that it can be a possible transparent conducting oxide (TCO)\cite{HJKim2,Ginley,Freeman,Minami,Hosono} and ignited even more studies. 

On the other hand, the importance of BSO as a transparent oxide semiconductor (TOS) has been noted and extensively studied.\cite{ChulkwonPark,UseongKim,Juyeon Shin,YoungMoKim,Sanchela} Its high carrier mobility and transparency as well as stable oxygen stoichiometry make BSO a promising material for optoelectronic devices.

\begin{figure}[ht!]
\centering
\includegraphics[scale=.33]{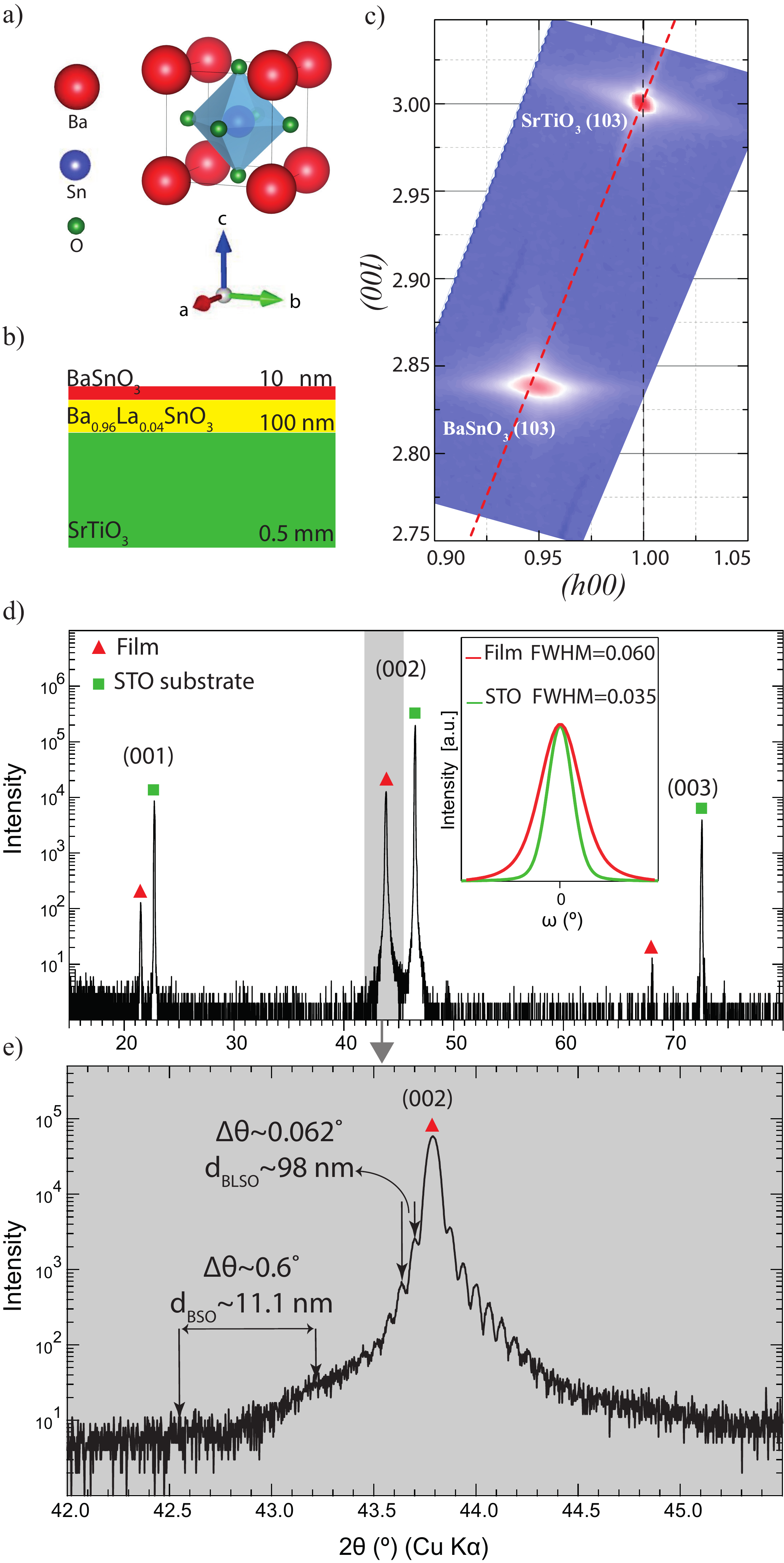}
\caption{(a) Perovskite cubic crystal structure of BaSnO$_3$ at room temperature. (b) Schematic of BaSnO$_3$/Ba$_{0.96}$La$_{0.04}$SnO$_3$/SrTiO$_3$ (10nm/100nm/0.5mm) thin-film grown by PLD method (Figure is not drawn to scale). (c) Reciprocal space mapping data of (103) peaks. Red dashed-line shows the symmetry axis of cubic crystal lattice. (d) X-ray diffraction 2$\theta$ scan shows corresponding peaks for BSO thin-film (red triangles) and STO substrate (green squares). The inset shows $\omega$-rocking curve data for the thin-film (red) and the substrate (green). FWHM for the film is 0.06 while that for the substrate is 0.035. (e) XRD data in the range $42-45.5^{\circ}$ clearly shows fringe pattern for the film. $\Delta\theta$ is the angular distance between fringe pattern peaks from which we estimate the thickness of the top BSO and BLSO buffer layers as 11.1 and 98 nm.}
\label{fig:figure1Methods}
\end{figure}

The importance of BSO as a TOS naturally calls for a need to study its physical properties. As the electronic and optical properties are mostly determined by the electronic structure, study of electronic structure, especially that of surface/interface, is one of the most essential properties that has to be done for BSO as an electronic material. For that reason, the electronic structure has been investigated by using first principles and tight-binding calculations\cite{BGKim,HRLiu,Moreira,Stanislavchuk,Scanlon,Sallis,Higgins,Krishnaswamy,HyunsuSim}. As for the experimental determination of the band structure, it has been limited to couple of angle-resolved photoemission spectroscopy (ARPES) studies on BSO\cite{BSJoo} and BLSO films\cite{Lochocki} which suggest for an indirect band gap\cite{BSJoo} and an upward band bending at the vacuum interface\cite{Lochocki}. However, the study on BSO was performed on a 8 nm thick film on SrTiO$_3$ (STO) substrate and possible distorted crystal structure from lattice mismatch between BSO film and STO substrate makes it unclear if the results represents the intrinsic property of BSO. It thus calls for additional electronic studies on relaxed films by ARPES to compare results with that of bulk crystals.

Surface reconstruction is an important issue for exploiting its potential for nano-scale electronic devices. For example, knowledge on the surface reconstruction of STO under various annealing conditions can provide important information for oxide electronics application using STO. The effect of oxygen vacancy (OV) on surface structure and in turn on the electronic structure is another important aspect of BSO as an electronic material candidate that also requires further study to previous works\cite{Mun,Iftekhar,Luo}. All these suggest for more surface studies on relaxed films by surface-sensitive techniques such as LEED and ARPES.

To have a better picture of surface atomic and electronic structures of BSO film, and to understand how those will react to different annealing conditions, we perform LEED and ARPES experiments on the surface of relaxed BSO thin-films. We combine LEED and ARPES measurements to study the effect of UHV and oxygen annealing on the surface and electronic structures and thermal stability of the film. The results show clear band dispersions which are partially consistent with those of a first principles calculation bulk\cite{Scanlon}. Our results not only provide a better understanding of the electronic structure of BSO thin-film but also reveal OV-induced reversible surface reconstruction after \emph{in-situ} UHV annealing. In addition, energy-momentum dispersion from ARPES measurement suggests that the valence band maximum (VBM) is located at 3.3 eV below Fermi level.

\begin{figure*}[ht!]
\centering
\includegraphics[scale=1]{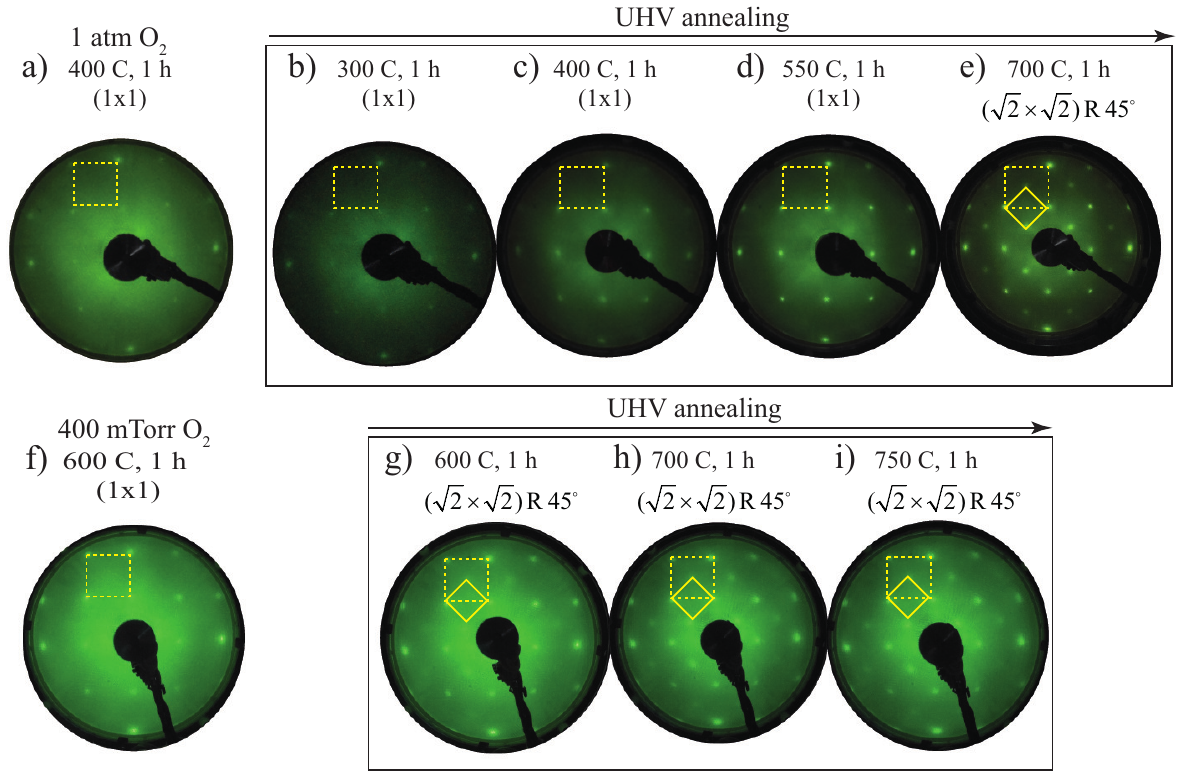}
\caption{90 eV LEED patterns of BaSnO$_3$ thin-film after annealed at (a) 400$^{\circ}$C and 1 atm oxygen with 1$\times$1 surface structure as shown by dashed yellow square. (b), (c), and (d) 300, 400, and 550$^{\circ}$C in UHV condition, respectively and (e) 700$^{\circ}$C; 1$\times$1 surface reconstructs into a $\sqrt{2}$$\times$$\sqrt{2}R45^\circ$ surface shown by yellow square. (f) After annealing at 600$^{\circ}$C in 400 mTorr oxygen; $\sqrt{2}$$\times$$\sqrt{2}R45^\circ$ disappears. (g), (h), and (i) 600, 700, and 750$^{\circ}$C in UHV condition.  $\sqrt{2}$$\times$$\sqrt{2}R45^\circ$ reconstruction is reproduced as a result of a second UHV annealing. Sample was not exposed to air after \emph{ex-situ} step in (f) (refer to Fig. 3 caption). All UHV annealing data shown in (b)-(e) and (g)-(i) were taken without exposure to air.}
\label{fig:figure2SurfaceStructure}
\end{figure*}

\section{Methods}

\subsection{Sample growth}

Thin-films of BaSnO$_3$ (BSO) were epitaxially grown using pulsed laser deposition (PLD) method from a polycrystalline target with cubic perovskite crystal structure and lattice constant of 4.116 $\mathrm{\AA}$ \cite{HJKim}  on top of a SrTiO$_3$ (STO) (001) surface at 750$^{\circ}$C (Fig. \ref{fig:figure1Methods}(a)). The lattice constant of BLSO is almost identical to that of BSO\cite{HJKim}. The base pressure in the PLD chamber was $1.4 \times 10^{-7}$ Torr. The partial oxygen pressure in the PLD chamber during growth was 100 mTorr. The growth condition was similar to that used for the work in Ref \cite{Mun}. KrF excimer laser (248 nm) with energy density of 1.43 $J/cm^{2}$ and intensity of 42.2 mJ was used. To grow BSO thin-film, first a 100 nm buffer layer of Ba$_{0.96}$La$_{0.04}$SnO$_3$ (BLSO) was grown on the STO substrate to compensate the lattice mismatch between STO substrate and BSO top layer. Then top-most BSO film was grown with a thickness of 10 nm (Fig. \ref{fig:figure1Methods}(b)).

\subsection{X-ray diffraction}
To check the epitaxial growth of thin-film we did X-ray diffraction (XRD), reciprocal space mapping (RSM) and rocking curve measurements (Fig. \ref{fig:figure1Methods}(c)-(d)). The X-ray measurement was performed using a D8 Discover high-resolution x-ray diffraction (Bruker) machine at Institute for Basic Science, center for correlated electron systems (IBS-CCES) located at Seoul National University (SNU). Figure \ref{fig:figure1Methods}(c) is the RSM data which shows the diffraction patterns for STO and BSO. From the symmetry axis of cubic crystal lattice (shown by red dashed-line in Fig. \ref{fig:figure1Methods} (c)) it is obvious that the BSO peak has approached the line suggesting that the lattice mismatch between STO substrate and film has dramatically decreased. Figure \ref{fig:figure1Methods}(d) shows $\theta$/2$\theta$ wide scan. Peaks from BSO thin-film are shown with red triangles and peaks from STO substrate are shown with green squares. Our calculation shows that the experimental values for in-plane and out-of-plane lattice constants of BSO thin film are 4.113 $\mathrm{\AA}$ and 4.130 $\mathrm{\AA}$, respectively. Comparison of these values with those of bulk BSO (4.116 $\mathrm{\AA}$) and STO substrate (3.906 $\mathrm{\AA}$) shows that our film is almost relaxed with slight compressive strain and elongation in the in- and out-of-plane directions, respectively. For a full characterization of the film including transmission electron microscopy (TEM), resistivity, and Hall mobility data, please see Ref \cite{Mun}. The inset image in Fig \ref{fig:figure1Methods}(d) shows fitted $\omega$ rocking curve data for the film and the substrate. Full width at half maximum (FWHM) for the film (red) is only 0.06 and 0.035 for the STO substrate (green), which indicates high quality epitaxial growth. Figure \ref{fig:figure1Methods}(e) is an enlargement of the XRD data in the 2$\theta$ range of $42^{\circ}$-$45.5^{\circ}$. The peak from the thin film (red triangle) shows clear fringe pattern which is a characteristic of thin-film XRD pattern. From angular distance between fitted fringe peaks in Fig. \ref{fig:figure1Methods}(e)), we estimate the thicknesses of BSO top layers (d$_{BSO}$) and BLSO buffer layers (d$_{BLSO}$) to be 11.1 and 98 nm, respectively.

\subsection{Surface preparation, LEED and ARPES}
BSO film samples were \emph{ex-situ} transferred from the PLD chamber to the preparation chamber equipped with a SPECS ErLEED instrument. The pressure in the preparation chamber was better than $2\times10^{-11}$ Torr. Surface preparation was done using electron-beam (e-beam) heating. The annealing temperature was monitored using a pyrometer with the emissivity of BSO thin-film set to 0.35. We performed series of UHV and oxygen annealing on the sample and take LEED data for each annealing temperature. 

\begin{figure*}[t]
\centering
\includegraphics[scale=.48]{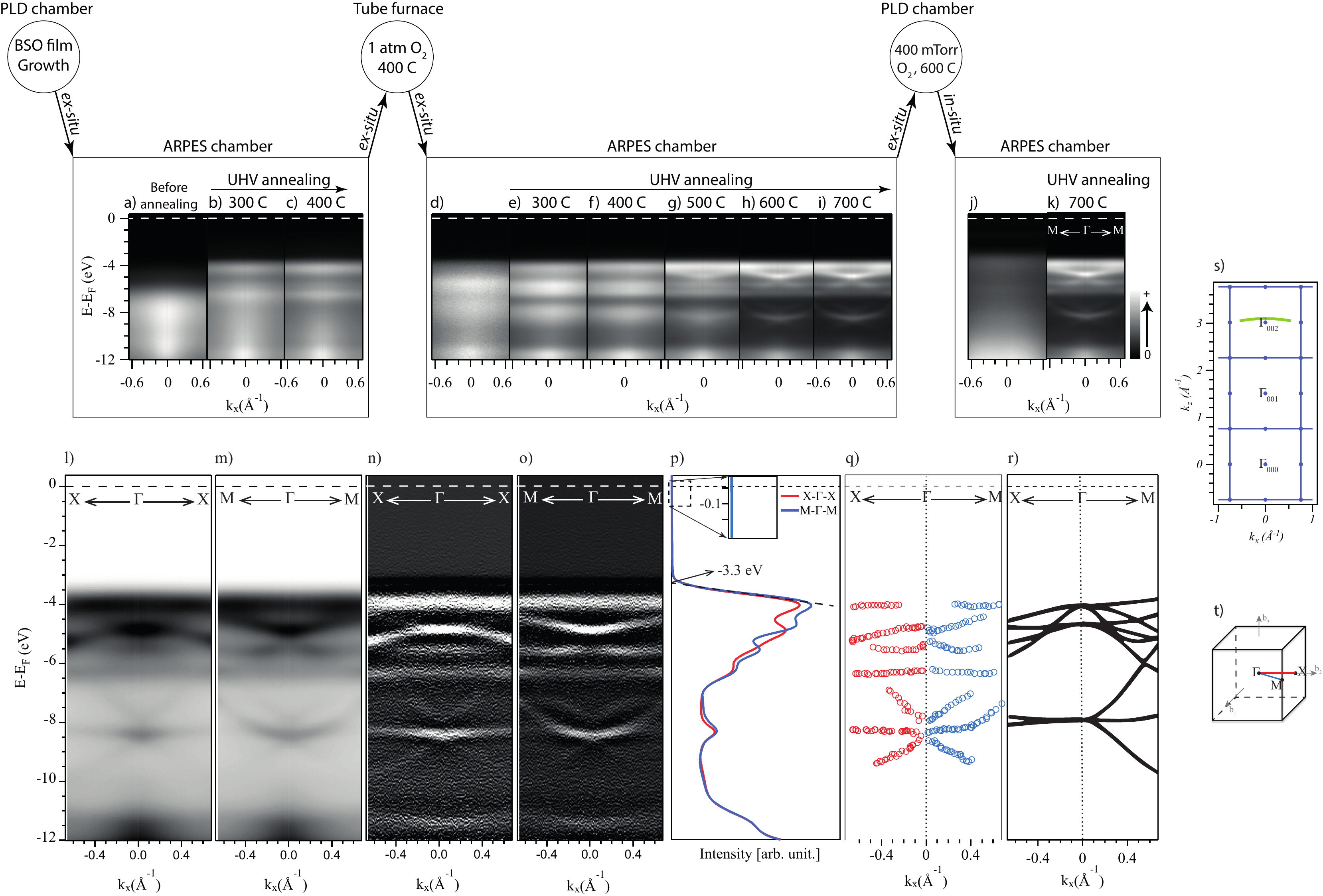}
\caption{BSO film grown in a PLD chamber is \emph{ex-situ} transferred to a UHV chamber for surface studies (LEED and ARPES). (a), (b), (c) ARPES data before annealing, after UHV annealing at 300$^{\circ}$C, and 400$^{\circ}$C. Afterwards, sample is \emph{ex-situ} transferred to a tube furnace to be annealed at 400$^{\circ}$C and 1 atm flowing oxygen. (d) ARPES data taken after sample is \emph{ex-situ} transferred back from tube furnace to the ARPES chamber. (e)-(i) Sample annealed in UHV condition starting from 300$^{\circ}$C up to 700$^{\circ}$C. Then, sample is \emph{ex-situ} transferred to PLD chamber to be annealed at 600$^{\circ}$C in 400 mTorr O$_2$. (j) ARPES data taken after sample was \emph{in-situ} transferred back from PLD chamber to the ARPES chamber. (k) ARPES after annealing at 700$^{\circ}$C in UHV. All ARPES data were taken with 21.2 eV photons at low temperature (7-12K). Electronic structure of BaSnO$_3$ thin-film taken along (l) X-$\Gamma$-X and (m) M-$\Gamma$-M. (n), (o) Second derivative of the data presented in (l) and (m), respectively. (p) Momentum integrated energy distribution curves (EDCs) for X-$\Gamma$-X (red) and M-$\Gamma$-M (blue) directions. The inset shows EDCs just below the Fermi level with no sign of in-gap states. Linear fit of leading edge shown by dashed line. (q) Extracted ARPES bands from momentum distribution curve (MDC) peak fitting. (r) DFT results for bulk reproduced from Ref. \cite{Scanlon} with permission from \text{\textcopyright}2013 American Physical Society. (s) Probed location in the k$_z$ direction in the Brillouin zone (BZ). (t) Cubic BZ of BaSnO$_3$. Red and blue lines show the high-symmetry cuts which our ARPES data were taken along. Samples were annealed in the preparation chamber at 750$^{\circ}$C for 1 hour before \emph{in-situ} transfer to the ARPES chamber. ARPES data were taken in a measurement temperature of 12 K and photon energy of 21.2 eV.}
\label{fig:figure3ElectronicStructure}
\end{figure*}

For each annealing and LEED measurement, the sample was \emph{in-situ} transferred to the ARPES chamber equipped with a DA30 Scienta Omicron hemispherical electron analyzer for electronic structure measurements. All data were taken at low temperature ($\sim$7-12 K) using a helium discharge lamp with photon energy of 21.2 eV. High symmetry cuts along X-$\Gamma$-X and M-$\Gamma$-M Data were taken near the $\Gamma$$_{002}$ point as shown in Fig. \ref{fig:figure3ElectronicStructure}(s) and (t). The pressure in the measurement chamber was better than $1.5\times10^{-10}$ Torr.

\section{results}
\subsection{Surface structure (LEED)}
Figure \ref{fig:figure2SurfaceStructure}(a) shows the LEED data from BSO film after \emph{ex-situ} post-annealing in a tube furnace in 1 atm oxygen pressure at 400$^{\circ}$C. A 1$\times$1 surface structure is visible from LEED diffraction spots as demonstrated by a yellow dashed square in Fig. \ref{fig:figure2SurfaceStructure}(a). We should mention here that the previous step in \emph{ex-situ} annealing in 1 atm O$_2$ is necessary towards recovering the dispersive energy bands. Without this step, the transparent sample eventually becomes opaque, an indication that there is too much OV. This suggests a delicate balance between oxygen vacancy and annealing condition should be achieved. Figure S1 (See Supplemental Material \cite{Supplemental Material}) shows difference between two samples (A and B), with and without step shown in Fig. \ref{fig:figure2SurfaceStructure}(f). Fig. S1 (a) and (b) shows that while sample A with \emph{ex-situ} O$_2$ annealing remains transparent even with UHV annealing at 700$^{\circ}$C, sample B starts to become opaque as we increase the UHV annealing temperature. Fig. S1 (c), (d) shows energy distribution curves (EDCs) taken at k$_x$=0 for two samples. For sample B, it can be seen that a broad signal starts to appear at 2.5 - 3 eV below the Fermi level (shown by blue arrow compared to sample A's shown by red arrow). We attribute the broad signal to possible oxygen deficient in-gap states which are expected as the 1 atm 400$^{\circ}$C O$_2$ annealing was not done for this sample.

Next, we post-anneal the BSO thin-film \emph{in-situ} in UHV condition starting from 300$^{\circ}$C up to 550$^{\circ}$C as shown in Fig. \ref{fig:figure2SurfaceStructure} (b)-(d). LEED pattern improves and we see sharper diffraction spots as we increase annealing temperature indicating that the surface is getting cleaner under UHV annealing. The 1$\times$1 surface structure is clearly seen (demonstrated by yellow dashed squares). Another UHV annealing at 700$^{\circ}$C, shows that new features appear (yellow square) indicating a surface reconstruction (Fig. \ref{fig:figure2SurfaceStructure}(e)) which will be discussed in details in the Discussion section as a new $\sqrt{2}$$\times$$\sqrt{2}R45^\circ$ surface reconstruction.

Annealing in 400 mTorr oxygen pressure at 600$^{\circ}$C, it is seen that the $\sqrt{2}$$\times$$\sqrt{2}R45^\circ$ surface reconstruction is reversed to 1$\times$1 surface (Fig. \ref{fig:figure2SurfaceStructure} (f)).  Another series of UHV annealing on the sample reproduces the $\sqrt{2}$$\times$$\sqrt{2}R45^\circ$ surface reconstruction. The results are shown in Fig. \ref{fig:figure2SurfaceStructure} (g)-(i). Figure \ref{fig:figure2SurfaceStructure} (g) shows that after UHV annealing at 600$^{\circ}$C the $\sqrt{2}$$\times$$\sqrt{2}R45^\circ$ surface reconstruction starts to appear again and at 700$^{\circ}$C it becomes more clear (Fig. \ref{fig:figure2SurfaceStructure} (h)), and is stable up to 750$^{\circ}$C annealing as shown in Fig. \ref{fig:figure2SurfaceStructure} (i). We should notice that the LEED pattern shown in Fig. \ref{fig:figure2SurfaceStructure} (e) is sharper than that of Fig. \ref{fig:figure2SurfaceStructure} (h) due to \emph{ex-situ} step of 400 mTorr oxygen annealing.

\subsection{Electronic structure (ARPES)}
In order to have deeper insight into the behavior of BSO surface as far as surface reconstruction is concerned, it is important to have information on the electronic structure. Figures \ref{fig:figure3ElectronicStructure} summarize our ARPES data from BSO thin-film. Figure \ref{fig:figure3ElectronicStructure} (a)-(k) shows post-annealing temperature dependent ARPES data in UHV/O$_2$ condition which will be discussed in details in next section. Data shown in Figs. \ref{fig:figure3ElectronicStructure} (l) and (m) are taken along X-$\Gamma$-X and M-$\Gamma$-M, respectively, measured at 12 K after \emph{in-situ} UHV annealing at 750$^{\circ}$C. Figs. \ref{fig:figure3ElectronicStructure} (n) and (o) show second derivatives of the ARPES data presented in panels (l) and (m). From a linear fit of the leading edge of integrated EDCs, the valence band maximum (VBM) is estimated to be at 3.3 eV below the Fermi level as shown in Fig.  \ref{fig:figure3ElectronicStructure} (p). Several dispersive bands are observed in the data. The peak locations of EDCs extracted from ARPES data in Figure \ref{fig:figure3ElectronicStructure} (l) and (m) are shown in Fig. \ref{fig:figure3ElectronicStructure} (q). Figure \ref{fig:figure3ElectronicStructure} (r) is DFT calculation for bulk re-produced from Ref. \cite{Scanlon}. Figure \ref{fig:figure3ElectronicStructure} (s),(t) show the probed momentum points in the BZ with 21.2 eV photons and the cubic BZ of BSO where ARPES data were taken along high symmetry cuts.

\section{Discussion}

In this section, we discuss our surface (LEED) and electronic (ARPES) structure data and how they reveal a new type of surface reconstruction for BSO film that has not been reported so far. We discuss the role of oxygen vacancies in the new surface reconstruction and its impact on stability of BSO surface.

Before we discuss the surface and electronic structures presented in the previous section, we wish to touch upon on the robustness of BSO film surface. As already described, PLD grown BSO thin-films were \emph{ex-situ} transferred to UHV chambers for surface studies. Our very first ARPES data of \emph{ex-situ} transferred film without any surface treatment already shows broad spectra as presented in Fig. \ref{fig:figure3ElectronicStructure} (a). Subsequent UHV annealing at 300$^{\circ}$C and 400$^{\circ}$C \emph{in-situ} led to better defined electronic structure as shown in Fig. \ref{fig:figure3ElectronicStructure} (b),(c). Non-dispersive bands are clearly seen in the ARPES data. We point out that observation of even non-dispersive bands are surprising for an \emph{ex-situ} transferred oxide system, implying that the surface must be quite stable. As we are more interested in dispersive bands, we did further annealing studies of the film: annealing in 1 atm O$_2$ in a tube furnace and subsequent annealing in UHV chamber. ARPES measurements show clear valence bands  as will be discussed below. Therefore, a key observation here is that BSO has rather a stable surface in comparison with other oxides, considering we are able to distinguish individual bands repeatedly for various annealing conditions.

Figure \ref{fig:figure2SurfaceStructure} (a)-(d) show surface structure results after 1 atm O$_2$ annealing and UHV annealing at 300$^{\circ}$C, 400$^{\circ}$C, and 550$^{\circ}$C, respectively. Data show a 1$\times$1 surface (dashed yellow square) suggesting that 1$\times$1 surface of BSO is stable under UHV annealing up to 550$^{\circ}$. The ARPES data taken after each annealing (Fig. \ref{fig:figure3ElectronicStructure} (d)-(h)) shows gradual change towards dispersive bands. 

The most interesting observation starts to appear when we increase the UHV annealing temperature to 700$^{\circ}$C. The LEED data suggest that the previous 1$\times$1 surface is now reconstructed into a $\sqrt{2}$$\times$$\sqrt{2}R45^\circ$ surface. The relevant reconstructed surface unit cell is shown in Fig. \ref{fig:figure2SurfaceStructure}(e) by the solid yellow square compared to the underlying 1x1 unreconstructed lattice unit cell (shown by dashed yellow square). This is the first evidence for such a surface reconstruction on BSO surface. As for the electronic structure, ARPES data taken right after LEED show that non- or weakly-dispersive energy bands from 1$\times$1 are now changed into dispersive bands for $\sqrt{2}$$\times$$\sqrt{2}R45^\circ$ surface (Fig \ref{fig:figure3ElectronicStructure} (i)). 

A natural question to ask is what the origin of the new surface reconstruction is. To answer this question, we should notice that it appeared as a result of consecutive UHV annealing and exposure to incident light from photon source. This observation suggests that OVs can be the culprit for the surface reconstruction. To test this claim, sample was transferred to a chamber and heated in 400 mTorr O$_2$ at 600$^{\circ}$C for 1 hour. Then, it was cooled down to room temperature and then \emph{in-situ} transferred to the UHV chamber where LEED experiment was performed. The result is presented in Fig. \ref{fig:figure2SurfaceStructure}(f), showing the $\sqrt{2}$$\times$$\sqrt{2}R45^\circ$ being reversed to a 1$\times$1 surface after oxygen annealing. This suggests OV induced by UHV annealing as the driving force for $\sqrt{2}$$\times$$\sqrt{2}R45^\circ$ surface reconstruction in the surface of BSO thin-film. As for the electronic structure, ARPES data show that the dispersive bands become non-dispersive again (Fig. \ref{fig:figure3ElectronicStructure} (j)).

To support the role of oxygen vacancies in $\sqrt{2}$$\times$$\sqrt{2}R45^\circ$ surface reconstruction even further, we performed another round of UHV annealing (to induce oxygen vacancies again), LEED and ARPES measurements starting from 600$^{\circ}$C. The idea is to see the effect of UHV annealing and consequent OVs on the surface, and also to see if the newly discovered surface reconstruction is reproduced by UHV annealing (after it had been oxygen annealed) or not. The result is presented in Fig. \ref{fig:figure2SurfaceStructure}(g) which shows that the $\sqrt{2}$$\times$$\sqrt{2}R45^\circ$ starts to reappear. Increasing the annealing temperature to 700$^{\circ}$C and 750$^{\circ}$C leads to even more clear $\sqrt{2}$$\times$$\sqrt{2}R45^\circ$ spots (Fig. \ref{fig:figure2SurfaceStructure} (h),(i)). As for the relevant ARPES data, the dispersive bands reappear too (Fig. \ref{fig:figure3ElectronicStructure} (k)). The emergence of new $\sqrt{2}$$\times$$\sqrt{2}R45^\circ$ surface reconstruction on BSO after UHV annealing, its reversal to 1$\times$1 after O$_2$ annealing, and its recovery after another UHV annealing provide conclusive evidences for OV driven $\sqrt{2}$$\times$$\sqrt{2}R45^\circ$ surface reconstruction in BSO thin-film.

Another aspect of our study is the relation between surface structure and the electronic structure in light of newly discovered $\sqrt{2}$$\times$$\sqrt{2}R45^\circ$ surface reconstruction. Surface structure dependent electronic structure is an important aspect of other oxide surfaces such as STO\cite{Ogawa,Tanaka}. In order to discuss this issue, we look at the energy-momentum dispersion maps in Fig. \ref{fig:figure3ElectronicStructure} more closely. First, we note that, in contrast to previous ARPES results\cite{BSJoo} which reported additional in-gap states due to intrinsic defects or OVs, no spectral weight in ARPES near Fermi level are observed (the inset in Fig. \ref{fig:figure3ElectronicStructure} (p)) as expected from relaxed and defect-free surface of our BSO film grown on a BLSO buffer layer. Our observation that induced carrier or in-gape state density is extremely low even after 750$^{\circ}$C UHV annealing, clearly showing the stability of the BSO surface. Other than the in-gap states, our ARPES data has some similarity to the earlier work\cite{BSJoo}.  It is also in overall agreement with calculated dispersions of BSO and BLSO\cite{Lochocki,Scanlon}. We have re-produced band structure calculation of Ref. \cite{Scanlon} to directly compare to our ARPES data as shown in Fig. \ref{fig:figure3ElectronicStructure} (r) . Any inconsistency between our band structure data from ARPES on the film surface with those of DFT calculation is not surprising as the calculation in Ref. \cite{Scanlon} has been done for a perfect bulk crystal while our ARPES data is from reconstructed surface of thin-film. Integrated EDCs for X-$\Gamma$-X and M-$\Gamma$-M in Fig. \ref{fig:figure3ElectronicStructure} (p) show peaks located at 4, 4.7, 5.5, 6.3, and 8.4 eV below the Fermi level, with the top of the band edge approximately at 3.3 eV while at least six dispersive bands are seen in band calculation results\cite{Lochocki,Scanlon}.

Perhaps, what is more important is the electronic structure difference between 1$\times$1 and $\sqrt{2}$$\times$$\sqrt{2}R45^\circ$ surfaces, and how they might relate to carrier mobilities in the surface and interface. While the 1$\times$1 surface shows non-dispersive bands, the $\sqrt{2}$$\times$$\sqrt{2}R45^\circ$ surface shows very clear dispersions. If these are intrinsic properties of the two surfaces, the mobility of the carriers in $\sqrt{2}$$\times$$\sqrt{2}R45^\circ$ surface are expected to be higher. In fact, increasing the mobility of charge carriers in BLSO thin-film has been studied and post-treatment methods such as vacuum annealing is believed to be a way to increase the mobility\cite{Shiogai,Cho,Yoon,Yu}. Post-treatment methods seems to be realistic and cost-effective for industrial production purposes. In that sense, our findings could provide useful information on how UHV/O$_2$ annealing affects the surface and electronic structures necessary to further works to improve the carrier motion at the interfaces and suggest how further transport studies on UHV/O$_2$ annealed BSO and BLSO surfaces should be done. We note that existing transport measurement results cannot say much about the conductance of the BSO $\sqrt{2}$$\times$$\sqrt{2}R45^\circ$ surface and how it compares with 1$\times$1 surface. Further study maybe necessary to address this issue.

\begin{table}[h]
\caption{Summary of UHV surface and electronic structures studies for BSO thin-film using LEED and ARPES. T: Annealing temperature. t: Annealing time (hour). P: Pressure, RC: Surface reconstruction. ARPES: dispersive ($\surd$) or non-dispersive (x) bands from ARPES measurement.}
\resizebox{\columnwidth}{!}{%
\begin{tabular}{c c c c c}\\
\hline\hline\\
 T ($^{\circ}$C) & t (h) & P & RC & ARPES\\ [0.5ex]
 \hline\\
Pre-annealing&---&UHV&(1$\times$1)&x\\
 \hline
 $\downarrow$ 300&1&UHV&(1$\times$1)&x\\
 \hline
 $\downarrow$ 400&1&UHV&(1$\times$1)&x\\
 \hline\\
 $\mapsto$ 400&1&1 atm O$_2$&(1$\times$1)&x\\
 \hline\\
 $\downarrow$ 300&1&UHV&(1$\times$1)&x\\
 \hline
 $\downarrow$ 400&1&UHV&(1$\times$1)&x\\
 \hline
 $\downarrow$ 500&1&UHV&(1$\times$1)&x\\
 \hline
 $\downarrow$ 700&1&UHV&$\sqrt{2}$$\times$$\sqrt{2}R45^\circ$&$\surd$\\
 \hline\\
 $\mapsto$ 600&1&400 mTorr O$_2$&(1$\times$1)&x\\
 \hline\\
 $\downarrow$ 600&1&UHV&$\sqrt{2}$$\times$$\sqrt{2}R45^\circ$&$\surd$\\
 \hline
 $\downarrow$ 700&1&UHV&$\sqrt{2}$$\times$$\sqrt{2}R45^\circ$&$\surd$\\
 \hline
 $\downarrow$ 750&1&UHV&$\sqrt{2}$$\times$$\sqrt{2}R45^\circ$&$\surd$\\
 \hline\hline
 \end{tabular}
}
 \label{table:summary}
 \end{table}

\section{Summary}

In summary, we performed LEED and ARPES experiments on PLD grown thin-films of BaSnO$_3$ to probe surface and electronic structures. Our data show that UHV annealing up to 550$^{\circ}$C gives 1$\times$1 LEED pattern with non-dispersive bands from ARPES data which suggest that BSO surface is stable at this temperature. Annealing at 700$^{\circ}$C leads to a new $\sqrt{2}$$\times$$\sqrt{2}R45^\circ$ surface reconstruction with dispersive bands from ARPES measurement and valence band edge located at 3.3 eV below the Fermi level. The $\sqrt{2}$$\times$$\sqrt{2}R45^\circ$ reconstruction is reversed to 1$\times$1 by annealing at 600$^{\circ}$C and 400 mTorr oxygen. This is suggestive of the role of OVs in the development of new surface reconstruction. A second UHV annealing starting from 600$^{\circ}$C shows that the $\sqrt{2}$$\times$$\sqrt{2}R45^\circ$ appears again and remains the same for the annealing temperature range of 600-750$^{\circ}$C. As for electronic structure data, while our experimental data is in partial agreement with existing calculation results for bulk, further theoretical studies with both surface reconstructions considered should be done to address the remaining issues. Finally, we have summarized all our findings in Table. \ref{table:summary}

\section*{Acknowledgment}

We are thankful for fruitful discussions with Young Jun Chang. This work was supported by IBS-009-D1 and IBS-009-G2.

\renewcommand{\figurename}{Fig. S}
\setcounter{figure}{0}


\begin{figure*}[b]
\centering
\includegraphics[scale=.55]{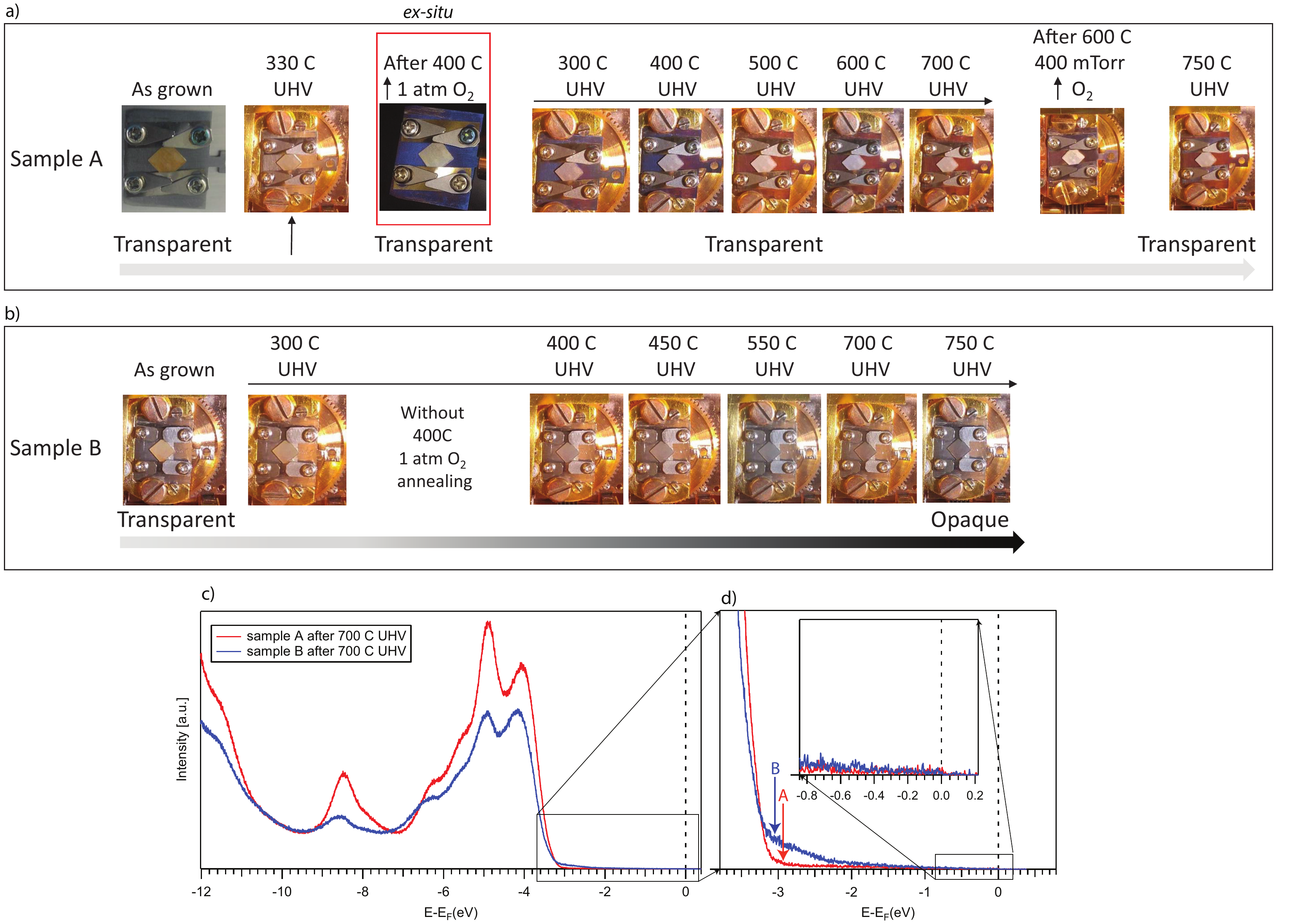}
\caption{Effect of 400$^{\circ}$C 1 atm O$_2$ \emph{ex-situ} annealing on transparency of the film. (a) Sample A with 400$^{\circ}$C 1 atm O$_2$ \emph{ex-situ}; As-grown film is transparent. Upon UHV annelaing at 300$^{\circ}$C it is still transparent and will stay up to 750$^{\circ}$C UHV annelaing. (b) sample B without 400$^{\circ}$C 1 atm O$_2$ \emph{ex-situ} annealing; As-grown transaprent sample goes under consequent UHV annealing starting from 300$^{\circ}$C up to 750$^{\circ}$C. Transparent samples grdually changes into an opaque one as the annelaing temperature is increased. (c),(d) Show energy distribution curves (EDCs) at k$_x$=0 for sample A and B. Blue curve for sample B shows emergence of a broad signal at 2.5-3 eV below Fermi level (blue arrow) while sample A has no such feature (red arrow). Inset shows no conduction band signal near Fermi level.}
\label{fig:S1}
\end{figure*}


\begin{thebibliography}{40}

\bibitem{Shan} C. Shan, T. Huang, J. Zhang, M. Han, Y. Li, Z. Hu, J. Chu, \textcolor[rgb]{0.00,0.07,1.00}{J. Phys. Chem. C, \textbf{118} 6994-7001, (2014)}.

\bibitem{Ismail-Beigi} Sohrab Ismail-Beigi, Frederick J. Walker, Sang-Wook Cheong, Karin M. Rabe, and Charles H. Ahn, \textcolor[rgb]{0.00,0.07,1.00}{APL Materails \textbf{3}, 062510 (2015)}.

\bibitem{SMXing} S.M. Xing, C. Shan, K. Jiang, J.J. Zhu, Y.W. Li, Z.G. Hu, J.H. Chu, \textcolor[rgb]{0.00,0.07,1.00}{J. Appl. Phys. \textbf{117}, 103107, (2015)}.

\bibitem{WJ LEE} Woong-Jhae Lee, Hyung Joon Kim, Jeonghun Kang, Dong Hyung Jang, Tai Hoon Kim, Jeong Hyuk Lee, and kee Hoon Kim, \textcolor[rgb]{0.00,0.07,1.00}{Annual Review of Material Research \textbf{47}, 391-423 (2017)}.

\bibitem{Gogova} D. Gogova, A. Suwardi, Y.A. Kutznetsova, A.F. Zatsepin, L.A. Mochalov, A. Nezhdanov, B. Szyszka, \textcolor[rgb]{0.00,0.07,1.00}{International Journal of Advanced Applied Physics Reserach \textbf{4}, 1-8 (2017)}.

\bibitem{YajunZhang} Yajun Zhang, M. P. K. Sahoo and Jie Wang, \textcolor[rgb]{0.00,0.07,1.00}{Phys.Chem.Chem.Phys.,
\textbf{19}, 7032, (2017)}.

\bibitem{XLuo} X. Luo, Y.S. Oh, A. Sirenko, P. Gao, T.A. Tyson, K. Char, S.-W. Cheong, \textcolor[rgb]{0.00,0.07,1.00}{Appl. Phys. Lett. \textbf{100}, 172112, (2012)}.

\bibitem{Prakash} Abhinav Prakash, Peng Xu, Alireza Faghaninia, Sudhanshu Shukla, Joel W. Ager III, Cynthia S. Lo, and Bharat Jalan, \textcolor[rgb]{0.00,0.07,1.00}{Nature Communications \textbf{8}, Article number: 15167 (2017)}.


\bibitem{HJKim} Hyung Joon Kim, Useong Kim, Hoon Min Kim, Tai Hoon Kim, Hyo Sik Mun, Byung-Gu Jeon,
Kwang Taek Hong, Woong-Jhae Lee, Chanjong Ju, Kee Hoon Kim, and Kookrin Char, \textcolor[rgb]{0.00,0.07,1.00}{Applied Physics Express \textbf{5}, 061102 (2012)}.

\bibitem{Ginley} D. S. Ginley and C. Bright, \textcolor[rgb]{0.00,0.07,1.00}{MRS Bull. \textbf{25}, 15 (2000)}.

\bibitem{Freeman} A. J. Freeman, K. R. Poeppelmeier, T. O. Mason, R. P. H. Chang,
and T. J. Marks, \textcolor[rgb]{0.00,0.07,1.00}{MRS Bull. \textbf{25}, 45 (2000)}.

\bibitem{Minami} T. Minami, \textcolor[rgb]{0.00,0.07,1.00}{MRS Bull. \textbf{25}, 38 (2000)}.

\bibitem{Hosono} H. Hosono, \textcolor[rgb]{0.00,0.07,1.00}{Thin Solid Films \textbf{515}, 6000 (2007)}.

\bibitem{HJKim2} H.J. Kim, U. Kim, T.H. Kim, J. Kim, H.M. Kim, B.G. Jeon, W.J. Lee, H.S. Mun,
K.T. Hong, J. Yu, K. Char, K.H. Kim,\textcolor[rgb]{0.00,0.07,1.00}{ Phys. Rev. B. \textbf{86}, 165205, (2012)}.

\bibitem{ChulkwonPark} Chulkwon Park, Useong kim, Chan Jong Ju, ji sung park, Young Mo Kim, and kookrin Char, \textcolor[rgb]{0.00,0.07,1.00}{Applied Physics Letters \textbf{105}, 203503 (2014)}.

\bibitem{UseongKim} Useong Kim, Chulkwon Park, Taewoo Ha, Young Mo Kim, Namwook Kim, Chanjong Ju, Jisung Park, Jaejun Yu, Jae Hoon Kim, and Kookrin Char, \textcolor[rgb]{0.00,0.07,1.00}{APL Materials \textbf{3}, 036101 (2015)}.

\bibitem{Juyeon Shin} Juyeon Shin, Young Mo Kim, Youjung Kim, Chulkwon Park, and Kookrin Char, \textcolor[rgb]{0.00,0.07,1.00}{Applied Physics Letters \textbf{109}, 262102 (2016)}.

\bibitem{YoungMoKim} Young Mo Kim, Chulkwon Park, Useong Kim, Chanjong Ju, and Kookrin Char, \textcolor[rgb]{0.00,0.07,1.00}{Applied Physics Express \textbf{9}, 011201 (2016)}.

\bibitem{Sanchela} Anup V. Sanchela, Mian Wei, Joonhyuk Lee, Gowoon Kim, Hyoungjeen Jeen, Bin Feng, Yuichi Ikuhara, Hai Jun Choab and Hiromichi Ohta, \textcolor[rgb]{0.00,0.07,1.00}{J. Chem. C, \textbf{7}, 5797 (2019)}. 





\bibitem{BGKim} B.G. Kim, J.Y. Jo, S.W. Cheong, \textcolor[rgb]{0.00,0.07,1.00}{J. Solid State Chem. \textbf{197}, 134-138, (2013)}.

\bibitem{HRLiu} H.R. Liu, J.H. Yang, H.J. Xiang, X.G. Gong, S.H. Wei, \textcolor[rgb]{0.00,0.07,1.00}{Appl. Phys. Lett. \textbf{102}, 112109, (2013)}.

\bibitem{Moreira} E. Moreira, J.M. Henriques, D.L. Azevedo, E.W.S. Caetano, V.N. Freire,
E.L. Albuquerque, \textcolor[rgb]{0.00,0.07,1.00}{J. Solid State Chem. \textbf{187}, 186-194, (2012)}.

\bibitem{Stanislavchuk} T.N. Stanislavchuk, A.A. Sirenko, A.P. Litvinchuk, X. Luo, S.-W. Cheong, \textcolor[rgb]{0.00,0.07,1.00}{J. Appl. Phys. 112, 044108, (2012)}.

\bibitem{Scanlon} D.O. Scanlon, \textcolor[rgb]{0.00,0.07,1.00}{Phys. Rev. B. \textbf{87}, 161201, (2013)}.

\bibitem{Sallis} S. Sallis, D.O. Scanlon, S.C. Chae, N.F. Quackenbush, D.A. Fischer, J.C. Woicik, J.-
H.H. Guo, S.W. Cheong, L.F.J. Piper, \textcolor[rgb]{0.00,0.07,1.00}{Appl. Phys. Lett. \textbf{103}, 42105, (2013)}.

\bibitem{Higgins} Z. Lebens-Higgins, D. O. Scanlon, H. Paik, S. Sallis, Y. Nie, M. Uchida, N. F. Quackenbush, M. J. Wahila, G. E. Sterbinsky, Dario A. Arena, J. C. Woicik, D. G. Schlom, and L. F. J. Piper, \textcolor[rgb]{0.00,0.07,1.00}{Phys. Rev. Lett. \textbf{116}, 027602 (2016)}.








\bibitem{HyunsuSim} Hyunsu Sim, S. W. Cheong, and Bog G. Kim, \textcolor[rgb]{0.00,0.07,1.00}{Phys. Rev. B. \textbf{88}, 014101 (2013)}.

\bibitem{Krishnaswamy} Karthik Krishnaswamy, Burak Himmetoglu, Youngho Kang, Anderson Janotti, and Chris G. Van de Walle, \textcolor[rgb]{0.00,0.07,1.00}{Phys. Rev. B. \textbf{95}, 205202, (2017)}.




\bibitem{BSJoo} Beom Soo Joo, Young Jun Chang, Luca Moreschini, Aaron Bostwick, Eli Rotenberg, Moonsup Han, \textcolor[rgb]{0.00,0.07,1.00}{Current Applied Physics, \textbf{17}, 595-599, (2017)}.

\bibitem{Lochocki} Edward B. Lochocki, Hanjong Paik, Masaki Uchida, Darell G. Schlom, and Kyle M. Shen, \textcolor[rgb]{0.00,0.07,1.00}{Appl. Phys. Lett. \textbf{112}, 181603, (2018)}.



\bibitem{Mun} Hyosik Mun, Useong Kim, Hoon Min Kim, Chulkwon Park, Tai Hoon Kim, Hyung Joon Kim, Kee Hoon Kim, and Kookrin Char, \textcolor[rgb]{0.00,0.07,1.00}{Appl. Phys. lett. \textbf{102}, 252105 (2013).}

\bibitem{Iftekhar} H. M. Iftekhar Jaim, Seunghun Lee, Xiaohang Zhang, and Ichiro Takeuchi, \textcolor[rgb]{0.00,0.07,1.00}{Applied Physics Letters, \textbf{111}, 172102 (2017)}.

\bibitem{Luo} Bingcheng Luo and Junbiao Hu, \textcolor[rgb]{0.00,0.07,1.00}{ACS Appli. electron. \textbf{1}, 51 (2019).}




\bibitem{Supplemental Material} See Supplemental Material at [www.]

\bibitem{Ogawa} Shohei Ogawa, Koichi Kato, Naoki Nagatsuka, Shohei Ogura, and Katsuyuki Fukutani, \textcolor[rgb]{0.00,0.07,1.00}{Phys. Rev. B. \textbf{96}, 085303, (2017).}

\bibitem{Tanaka} Tomoaki Tanaka, Kenta Akiyama, Ryo Yoshino, and Toru Hirahara, \textcolor[rgb]{0.00,0.07,1.00}{Phys. Rev. B. \textbf{98}.  121410(R), (2018).}


\bibitem{Shiogai} Junichi Shiogai, Kazuki Nishihara, Kazuhisa Sato, and Atsushi Tsukazaki, \textcolor[rgb]{0.00,0.07,1.00}{AIP Advances \textbf{6}, 065305 (2016).}


\bibitem{Yoon} Daseob Yoon, Sangbae Yu, and Junwoo Son, \textcolor[rgb]{0.00,0.07,1.00}{NPG Asia Materials \textbf{10}: 363–371, (2018).}

\bibitem{Cho} Hai Jun Cho , Takaki Onozato, Mian Wei, Anup Sanchela, and Hiromichi Ohta, \textcolor[rgb]{0.00,0.07,1.00}{APL Mater. \textbf{7}, 022507 (2019).}

\bibitem{Yu} Sangbae Yu, Daseob Yoon, and Junwoo Son, \textcolor[rgb]{0.00,0.07,1.00}{Appl. Phys. Lett. \textbf{108}, 262101 (2016).}


\end{thebibliography}
\end{document}